# Electronic structure of oxygen-free 38K superconductor Ba$_{1-x}$K$_x$Fe$_2$As$_2$ in comparison with BaFe$_2$As$_2$ from first principles

I.R. Shein,* A.L. Ivanovskii

*Institute of Solid State Chemistry, Ural Branch of the Russian Academy of Sciences, Ekaterinburg GSP-145, 620041, Russia*

(Dated 04 June 2008)

**Based on first-principles FLAPW-GGA calculation, we have investigated electronic structure of newly discovered oxygen-free 38K superconductor Ba$_{1-x}$K$_x$Fe$_2$As$_2$ in comparison with parent phase - tetragonal ternary iron arsenide BaFe$_2$As$_2$. The density of states, magnetic properties, near-Fermi bands compositions, together with Sommerfeld coefficients γ and molar Pauli paramagnetic susceptibility χ are evaluated. The results allow us to classify these systems as *quasi-two-dimensional ionic metals*, where the conduction is strongly anisotropic, only happening on the (Fe-As) layers. According to our calculations, at the hole doping of BaFe$_2$As$_2$ the density of states at the Fermi level grows, and this can be a possible factor of occurrence of superconductivity for Ba$_{1-x}$K$_x$Fe$_2$As$_2$. On the other hand, Ba$_{1-x}$K$_x$Fe$_2$As$_2$ lays at the border of magnetic instability and the pairing interactions might involve magnetic or orbital fluctuations.**

Since the discovery (in February 2008 [1]) of superconductivity with $T_C \sim 26$K in the fluorine-doped quaternary La-Fe oxyarsenide LaO$_{1-x}$F$_x$FeAs, further promising developments [2,3] in search of related oxypnictide-based superconductors (SCs) are achieved, in particular with replacing La atoms by other rare-earth metals (*Ln* = Gd [4], Ce [5], Sm [6], Pr and Nd [7,8]), result in $T_C \sim$ 41-55K. Moreover, the comparable values of $T_C$ have been reported at replacing of rare-earth atoms by thorium (Gd$_{1-x}$Th$_x$OFeAs) [9] or Sr (La$_{1-x}$Sr$_x$OFeAs, where x ~ 0.09 - 0.20) [10] as well as for oxygen-deficient samples *Ln*O$_{1-\delta}$FeAs (*Ln* = Sm, Nd, Pr, Ce, La) without their doping [11].

---

* E-mail: shein@ihim.uran.ru

These unusual materials attract now considerable interest because are the first non-Cu-based layered superconductors adopting comparably high critical temperature and upper critical field, are located on border of magnetic instability, and should have an unconventional mechanism of superconductivity, which may be connected with magnetic fluctuations and a spin density wave (SDW) anomaly.

The parent phase - oxyarsenide LaOFeAs adopts the tetragonal ZrCuSiAs-type structure (space group P4/*nmm*) [12], where the positively charged (La-O)$^+$ layers alternate with the negatively charged (Fe-As)$^-$ layers along the *c* axis; the bonding between these layers is mostly ionic. On the other hand, though the chemical bonding in (La-O) layers is also ionic, the strong covalent interactions inside (Fe-As) layers occur. According to the available experimental and theoretical data [13-19] for LaOFeAs the electronic bands in a window around the Fermi level are formed mainly by states of (Fe-As) layers, whereas the bands of (La-O) layers are rather far from the Fermi level. Thus the superconductivity in LaFeAsO is determined mainly by structural and electronic states of the (Fe-As) layers.

Recently the related system: ternary iron arsenide $BaFe_2As_2$ was investigated and the SDW anomaly at 140 K similar to LaFeAsO was found [20]. Moreover this phase with the tetragonal $ThCr_2Si_2$-type structure (space group I4/*mmm*) [21] contains identical (Fe-As) layers formed from [$FeAs_4$] tetrahedrons which are separated by barium sheets instead of (La-O) layers for LaFeAsO. In both phases one electron is transferred to (Fe-As) layer according to ionic picture: $Ba^{2+} \rightarrow$ 0.5(FeAs)$^-$ and (LaO)$^+ \rightarrow$ (FeAs)$^-$. The authors [20] have assumed that $BaFe_2As_2$ may be suitable parent phase for search of new family of oxygen-free SCs. Really, quite recently [22] a superconducting transition at $T_C$ ~3 K for $BaNi_2P_2$, which belongs to the same $ThCr_2Si_2$-like structural type, has been fixed. Moreover, it was found [23] that the ternary iron arsenide $BaFe_2As_2$ with partial substitution K → Ba becomes superconducting up to $T_C$ ~ 38K, for $Ba_{0.6}K_{0.4}Fe_2As_2$.

In this Communications we present the results of the first-principles calculations of the electronic properties for above mentioned iron arsenide systems: namely for newly discovered 38K oxygen-free superconductor $Ba_{1-x}K_xFe_2As_2$ in comparison with parent phase - tetragonal ternary iron arsenide $BaFe_2As_2$.

The positions of the atoms for tetragonal arsenide $BaFe_2As_2$ are: Ba at 2*a* (0, 0, 0), Fe at 4*d* (0.5, 0, 0.25) and As at 4*c* (0, 0, *z*). To explore the superconducting phase, by the replacement of one barium atom in the double (a × a × c) unit cell of $BaFe_2As_2$ by a K atom we have simulated the hole-doped system with nominal composition $Ba_{0.5}K_{0.5}Fe_2As_2$ and space group C*mmm*, see Fig. 1.

Our band-structure calculations were carried out by means of the full-potential method with mixed basis APW+lo (LAPW) implemented in the WIEN2k suite of programs [24]. The generalized gradient correction (GGA) to exchange-correlation potential in the PBE form [25] was applied. The experimentally determined lattice parameters as well as internal positions *z* for $BaFe_2As_2$ and $Ba_{0.5}K_{0.5}Fe_2As_2$ (listed in Table 1) were used. Two series of calculations were performed: for the nonmagnetic (NM) and magnetic states - in approximation of FM ordering.

Firstly, for both LaOFeAs and $BaFe_2As_2$ the energy difference of magnetic state *versus* paramagnetic is very small and does not exceed at about 0.076 eV/form.unit for $BaFe_2As_2$ and at about 0.046 eV/form.unit for $Ba_{0.5}K_{0.5}Fe_2As_2$, *i.e.* these materials lays at the border of their magnetic instability.

Let us discuss the main peculiarities of the electronic structure of $BaFe_2As_2$ *versus* $Ba_{0.5}K_{0.5}Fe_2As_2$ using paramagnetic densities of states (DOSs) as depicted in Fig. 2.

For $BaFe_2As_2$, at high binding energies the quasi-core DOSs peaks are placed: from -14.3 eV to -12.9 eV with Ba 5*p* states, and from -12.0 eV to -10.2 eV with As 4*s* states, and also with some admixtures of Fe 3*d* and Ba 5*p* states. The valence band (VB) extends from -5.4 eV up to the Fermi level $E_F$ = 0 eV and includes three main subbands *A-C*, Fig. 2. Among them the first subband *A* ranging from the VB bottom up to -3.8 eV is predominantly formed from comparable contributions of As 4*p* and Fe 3*d* states. The next subband *B* (in region from -3.8 eV up to -2.1 eV) contains the main contributions from Fe 3*d* states, together with some admixture from As 4*p* states. Thus, the above mentioned subbands *A* and *B* are derived from the Fe 3*d* states hybridized with the As 4*p* states and are responsible for the covalent Fe-As bonding. Finally, the top of VB (subband *C*, in the interval from -2.1 eV up to $E_F$) is derived basically from the Fe 3*d* states. This Fe 3*d*-like band intersected with the Fermi level and continues up to +2.0 eV (unoccupied subband *D*); *i.e.* the near-Fermi region for $BaFe_2As_2$ is composed mainly by iron states with very small admixtures of As states.

Also, it is noticeable that the contributions from the valence states of barium into the occupied subbands *A-C* as well as into the bottom of the conduction subband *D* are negligible, *i.e.* in $BaFe_2As_2$ these atoms are in the form of the cations $Ba^{2+}$. This means that the Ba sheets and (Fe-As) layers are linked exclusively by a ionic interaction - as against quaternary oxyarsenide LaOFeAs, where between (La-O) and (Fe-As) layers some covalent bonding arises due to the partial overlapping of the La and As states, see for example [15-18]. Thus, our results indicate that iron arsenide $BaFe_2As_2$ consists of alternatingly stacked insulating $Ba^{2+}$ sheets and conductive $(Fe-As)^{1-}$ layers, and the bonding between them is of ionic type, *i.e.* this system may be described as *quasi-two-dimensional ionic metal*, where the conduction is strongly anisotropic, only happening on the (Fe-As) layers.

In turn, the overall shape of the valence DOS for $BaFe_2As_2$ is very similar to the potassium doped system, except the new sharp quasi-core peak near -14.3 eV which is derived from K 3*p* states and narrowing of Ba 5*p* peak, see Fig. 2. In both compounds, the valence band which extends from -5.4 eV up to the Fermi level (for BaFe$_2$As$_2$) and from -4.9 up to $E_F$ (for Ba$_{0.5}$K$_{0.5}$Fe$_2$As$_2$) is derived basically from the Fe 3$d$ states hybridized in the bottom of the VB with the As 4$p$ states; some distinctions of DOSs profiles between BaFe$_2$As$_2$ and Ba$_{0.5}$K$_{0.5}$Fe$_2$As$_2$ (see Fig. 2) are related to deformations of [FeAs$_4$] tetrahedrons building the (Fe-As) layers - by the partial replacement of barium by potassium, Table 1. In addition, for Ba$_{0.5}$K$_{0.5}$Fe$_2$As$_2$ the admixtures of Ba and K states in the VB are absent, *i.e.* this system preserved the type of *quasi-two-dimensional ionic metal*.

The most remarkable difference of the DOS for BaFe$_2$As$_2$ *versus* Ba$_{0.5}$K$_{0.5}$Fe$_2$As$_2$ is the location of the Fermi level, see Fig. 2. For hole-doped Ba$_{0.5}$K$_{0.5}$Fe$_2$As$_2$ the decrease of the band filling leads to the movement of the Fermi level in the region of higher binding energies. As result, $E_F$ in Ba$_{0.5}$K$_{0.5}$Fe$_2$As$_2$ is shifted downwards and is located on a slope of sharp peak $C$ in the region of enhanced DOS. Thus, the total density of states at the Fermi level N$(E_F)$ for Ba$_{0.5}$K$_{0.5}$Fe$_2$As$_2$ became at about 20% higher than the value of N$(E_F)$ for BaFe$_2$As$_2$, Table 2. Note that the growth of N$(E_F)$ was achieved exclusively due to Fe 3$d$ states, whereas the contribution from As states has remained less than 4 %, Table 2.

These data allow us also to estimate the Sommerfeld constants ($\gamma$) and the Pauli paramagnetic susceptibility ($\chi$) for iron arsenides BaFe$_2$As$_2$ and Ba$_{0.5}$K$_{0.5}$Fe$_2$As$_2$, assuming the free electron model, as: $\gamma = (\pi^2/3)N(E_F)k^2_B$, and $\chi = \mu_B^2 N(E_F)$. It is seen (Table 2) that both $\gamma$ and $\chi$ increase approximately at 20% as going from BaFe$_2$As$_2$ to Ba$_{0.5}$K$_{0.5}$Fe$_2$As$_2$; and these values are comparable with those as obtained for Fe-containing oxypnictides (for example $\gamma$ = 12.5 mJ·K$^{-2}$·mol$^{-1}$ for LaOFeP [28]).

Additionally the calculated Sommerfeld constant $\gamma^{theor}$ may be useful for simple estimations [29] of the average electron-phonon coupling constant $\lambda$ (for BaFe$_2$As$_2$-based SCs, in assumption of conventional BCS phonon-mediated mechanism of superconductivity) as $\gamma^{exp} = \gamma^{theor}(1 + \lambda)$. Within a very crude estimate, using our $\gamma^{theor}$ for BaFe$_2$As$_2$ (Table 2) and the measured [20] $\gamma^{exp}$ (~16 mJ/(mol·K$^2$)) we obtain an empirical value of λ about 0.5. *i.e.* these materials should be within the weak coupling limit. For comparison, for other superconducting oxygen-free species with magnetic ions, such as *A*CNi$_3$ anti-perovskites, the available experimental and theoretical estimations of λ varies from 1.4 to 0.66, see [30].

On the other hand, the superconducting hole-doped iron arsenide Ba$_{0.5}$K$_{0.5}$Fe$_2$As$_2$ is very similar to the above mentioned layered oxypnictides, for which a set of models of unconventional type of superconductivity was proposed - where, for example, the pairing interactions might involve magnetic or orbital fluctuations [13-17]. Note, that our results indicate that both Ba$_{0.5}$K$_{0.5}$Fe$_2$As$_2$ and Ba$_{0.5}$K$_{0.5}$Fe$_2$As$_2$ are placed at the border of magnetic instability, and the calculated magnetic moments for their FM states are at about 1.90 μ$_B$/Fe for BaFe$_2$As$_2$ and at about 0.74 μ$_B$/Fe for Ba$_{0.5}$K$_{0.5}$Fe$_2$As$_2$.

In summary, we studied the electronic structure of newly discovered 38K oxygen-free superconductor Ba$_{1-x}$K$_x$Fe$_2$As$_2$ in comparison with parent phase - tetragonal ternary iron arsenide BaFe$_2$As$_2$.

The density functional theory predicts that BaFe$_2$As$_2$ may be described as *quasi-two-dimensional ionic metal*, this compound consist of insulating Ba sheets and conductive (Fe-As) layers; the bonding between them is ionic, whereas the conduction is strongly anisotropic, only happening on the (Fe-As) layers.

According to our calculations, the result of hole doping through barium ions substitution is the enhance of the density of states at the Fermi level at about on 20%. This can be a possible factor of occurrence of superconductivity for Ba$_{1-x}$K$_x$Fe$_2$As$_2$. On the other hand, this system lays at the border of magnetic instability, and the possible role of such hole doping is the suppressing the SDW anomaly for inducing superconductivity [23]. Thus, the further in-depth studies are necessary to understand the possible scenarios of superconducting coupling mechanisms for these systems, which may be of interest as a new material platform for further exploration of relationships between magnetism and superconductivity for oxygen-free SCs.

**Table 1**. The lattice parameters (*a* and *c*, in Å), internal coordinates (*z*), some interatomic distances (*d*, in Å) and bond angles As-Fe-As (in deg.) for $BaFe_2As_2$ and $Ba_{1-x}K_xFe_2As_2$ [20,23].

| phase/parameter | *a* | *c* | *z* | *d*(Ba-As) |
|---|---|---|---|---|
| $BaFe_2As_2$ | 3.9625 | 13.0168 | 0.3538 | 3.382 |
| $Ba_{1-x}K_xFe_2As_2$ | 3.9090 | 13.2122 | 0.3538 | 3.372 |
| phase/parameter | *d*(Fe-As) | *d*(Fe-Fe) | bond angles | |
| $BaFe_2As_2$ | 2.403 | 2.802 | 111.1×2; 108.7×4 | |
| $Ba_{1-x}K_xFe_2As_2$ | 2.396 | 2.770 | 109.9 | |

**Table 2**. Total $N^{tot}(E_F)$ and partial $N^l(E_F)$ densities of states at the Fermi level (in states/eV·form.unit), electronic heat capacity γ (in mJ·K$^{-2}$·mol$^{-1}$) and molar Pauli paramagnetic susceptibility χ (in $10^{-4}$ emu/mol) for $BaFe_2As_2$ and $Ba_{0.5}K_{0.5}Fe_2As_2$.

| Phase/parameter | $N^{Fed}(E_F)$ | $N^{As}(E_F)$ | $N^{tot}(E_F)$ | γ | χ |
|---|---|---|---|---|---|
| $BaFe_2As_2$ | 1.860 | 0.071 | 4.553 | 10.73 | 1.47 |
| $Ba_{0.5}K_{0.5}Fe_2As_2$ | 2.352 | 0.072 | 5.526 | 13.03 | 1.79 |

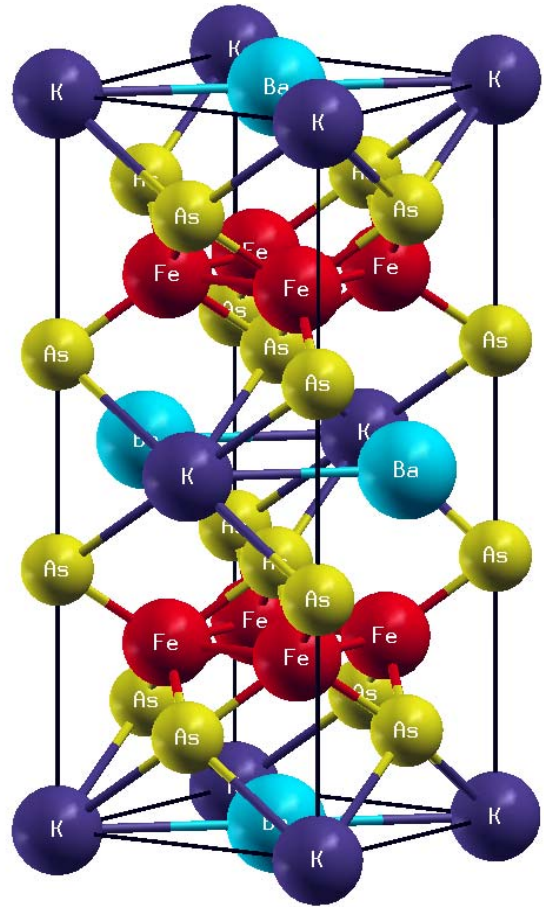

Figure. 1. Crystal structure of ordered phase $Ba_{0.5}K_{0.5}Fe_2As_2$.

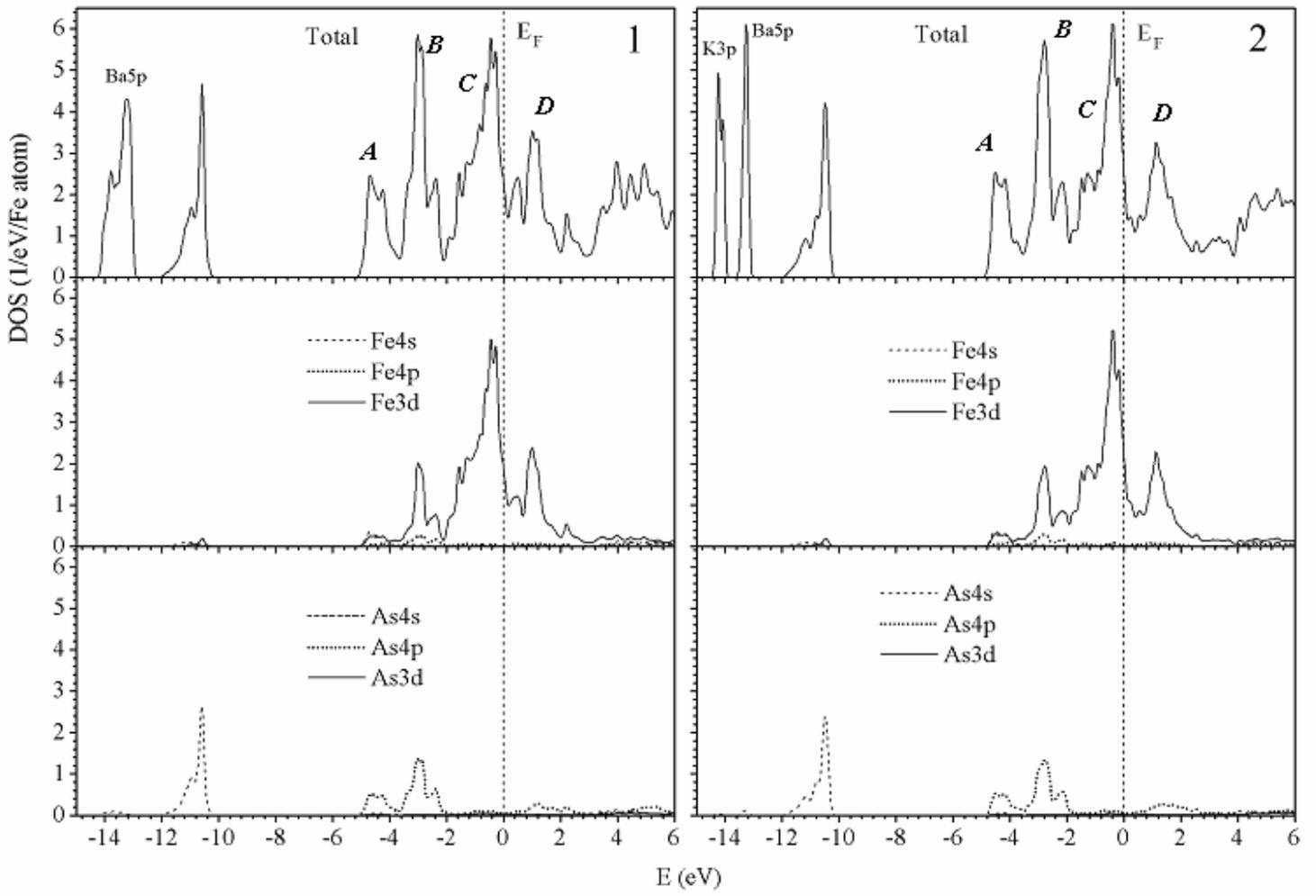

Figure 2. Total and partial densities of states for 1- $BaFe_2As_2$ (1) and $Ba_{0.5}K_{0.5}Fe_2As_2$ (2).